# Single Board Computers (SBC): The Future of Next Generation Pedagogies in Pakistan

Saad Wazir
Department of Computing
School of Electrical Engineering &
Computer Science,
National University of Sciences and
Technology (NUST), Islamabad, Pakistan
Preston University Islamabad, Pakistan
swazir.mscs18seecs@seecs.edu.pk

Hamza Ali Imran
Department of Computing
School of Electrical Engineering &
Computer Science,
National University of Sciences and
Technology (NUST), Islamabad, Pakistan
himran.mscs18seecs@seecs.edu.pk

Usama Latif
Operations Engineer
VAS, Apollo Telecom,
Islamabad
usamalatif417@yahoo.co.uk

Usama Mujahid
Department of embedded systems,
RwR Private Limited, I-9 /3, Islamabad,
Pakistan
umujahid363@gmail.com

Muhammad Bilal
Accelerated Software Team
Emumba Private Limited, I-10/3 I 10/3 I-10,
Islamabad, Pakistan
muhammad.bilal3967@gmail.com

*Abstract*—ARM processors have taken over the mobile industry from a long time now. Future of data centers and the IT industry is estimated to make use of ARM Processors. Projects like Openstack on ARM are enabling use of ARM in data centers . Single board computers (SBCs) based on ARM processors have become the norm these days. Reason for their popularity lies in their cost effective and power efficient nature. There are hundreds of them available in the market having different sizes, compute power and prices. The reason for their popularity is largely due to the rise of new technology called IoT (Internet of Things) but there is another perspective where they can become handy. Low Price and Power Usage of single board computers makes them top candidate to be used for teaching many courses with hands-on experience in developing countries like Pakistan. Many boards support full Linux distributions and can be used as general-purpose computers while many of them are open hardware based. In this paper, we have reviewed the famous options available and tried to figure out which of them are better for teaching what kind of courses.

I. INTRODUCTION

ARM-based single board computers (SBCs) are getting increasingly advanced and strong each day and beginning to compete with the x86-based systems of Intel, which for decades have been regarded as the best for personal computing. Not only are ARM systems cheap but because of their RISC architecture, they consume much less energy making them most appropriate for teaching in countries such as Pakistan. Almost all of these computers use various flavors of Linux which is the most famous open source operating system and is being used by all types of Computers ranging from the world's most powerful supercomputers to the tiny little embedded systems.

Desktop Linux has improved considerably since the last decade and now transition from Windows has started; which is also assumed as the only viable desktop operating system by a majority in our country. According to [1], Kerala an Indian state recently adopted Linux for school teaching by creating their own Linux distribution which is based on Ubuntu and also claimed it can save up to 428 Million US Dollars.

It is also believed that the ARM processors are going to replace the Intel Xeon processors in the future for even High Performance Computing. According to [2] ARM partners 100 billion chips, which makes it one of the most common architectures of its time. Since they are cheap and produce less heat, they can be used in abundance together. Projects like OpenStack on ARM are making that possible. Tools like Kubernetes, Message Passing Interface and Hadoop all run on ARM processors now. They are opening the doors of data centers for ARM processors. Therefore, computers having ARM processors and running open source software is the future of computing which makes them a better choice for teaching. There are many options available of SBCs in the market which makes selection of one a difficult task.

In this article, we will be comparing the most viable options available in the market and will go on to discuss in what terms SBCs can be used when it comes to teaching.

II. LITERATURE REVIEW

People have been using SBCs for teaching different areas of computing. Authors of [3] suggested making Raspberry Pi as a case study to teach computer architecture.

Use of Single Board Computers for teaching parallel computing has been suggested by many for some time now, traditionally virtual machines are used for creating clusters and teaching parallel and distributed computing but that does not give students hands on experience with actual hardware. Hence students remain unfamiliar with the problems which can arise when they work with real machines. [4] Doucet, K. suggested using Raspberry Pi 3b for teaching parallel and distributed computing. His student created a cluster of Raspberry Pi using MPICH 2, which is an open source implementation of Message Passing Interface and runs parallel simulation of Monte Carlo's Method for finding value of pi on that cluster.

A Raspberry Pi 3B cluster was implemented by authors of [5] to conduct experiments on two different Message Passing Interface implementations. The reason for using SBCs for such experiments was mainly because of their cost effectiveness. This illustrates SBCs have a potential for applications related to high performance computing and being cost effective can be used for giving students hands-on experience for cluster computing.

[6] presents an Introductory Information Systems course at The College of Engineering and Information Sciences at DeVry University making use of Raspberry Pi 2. This course gave students hands on experience with computers and programming.

N. A. M. Radzi et al [7] used similar kind of approach to teach "Programming for Engineers" course to undergraduate students at Universiti Tenaga Nasional, Malaysia. They integrate the learning of C programming language with BeagleBone Black and claimed that 80% of the students learned the desired skills.

Authors of [8] use Raspberry Pi to teach image and video processing by combining mathematical principles with engineering material for students using python libraries.

Nakano, Koji et al [9] uses FPGA board to teach the concepts of assembler, processor architecture, and compiler design. They concluded that students can better understand the CPU design, assembler design, Verilog HDL, and compiler design by doing hands-on with FPGA board.

Researchers [10] use a Raspberry Pi Zero to teach biology students basic computer networks and system administration to learn about their bioinformatic course. This eliminates the need for a specialized external lab for this task proving to be very cost effective.

H. A. Imran et al [11] have presented a naive model for providing a High Performance Compute Resource (HPC) as a service. Article is majorly focused on solving the problems faced by newcomers in the domain of HPC which includes the skills required to operate distributed systems. The presented solution solves that problem allowing the user to focus on programming the main task. Authors have presented a prototype which is made up of Raspberry Pi 3B and 3B+ SBCs. The presented prototype can be used for teaching Parallel & Distributed Computing, making use of SBCs.

Following Table (table 1) summary the above literature review

TABLE I. LITERATURE REVIEW SUMMARY

| Ref. | Year | Board | Area/Course |
|---|---|---|---|
| [3] | 2017 | Raspberry Pi | Computer Architecture |
| [4] | 2017 | Raspberry Pi | Parallel & Distributed Computing |
| [5] | 2019 | Raspberry Pi | Parallel & Distributed Computing |
| [6] | 2017 | Raspberry Pi | Information Systems |
| [7] | 2016 | BeagleBone Black | Programming for Engineers |
| [8] | 2017 | Raspberry Pi | Image and Video processing |
| [9] | 2009 | FPGA based board | Hardware Design |
| [10] | 2019 | Raspberry Pi | Computer Networks |
| [11] | 2019 | Raspberry Pi | Parallel & Distributed Computing |

### III. OVERVIEW OF BOARDS

In this section we have discussed the details of all the boards we believe are suitable for teaching different courses for Computer Science and Electrical or Computer Engineering programs.

*A. Raspberry pi 3B*

The most common SBC available in the market is Raspberry Pi (RPi) 3B shown in figure 1. It can be obtained easily from any local electronics store along with HDMI cable, power supply, Memory card, heat sink and a case accumulating its cost to about 45 US Dollars. [12] Broadcom's System-on-Chip (SoC) named BCM2837 which has 1.2 GHZ quad-core ARM Cortex A53 (ARMv8 Instruction Set), Broadcom VideoCore IV GPU clocked at 400 MHz, 1 GB of LPDDR2 RAM, 4 USB ports, 10/100 Mbps Ethernet port, 802.11n Wireless LAN and 4.0 Bluetooth. It does not have any storage available on board. Memory cards are used for storing all programs. It supports a lot of Linux distributions including Ubuntu, Fedora, Kali Linux etc, but the official OS by Raspberry pi Foundations recommended for it is named Raspbian. Its latest version is Buster. Raspbian is based on Debian which is one of the most famous Linux distribution. There are 3 Versions of Raspbian available, having a different number of tools available on them. The choice depends on one's needs. Raspbian. The new Raspbian Buster has 4.19 Kernel of Linux which is an LTS (Long Term Support) version. It is packaged with the educational tools including BlueJ Java IDE, Geany Programmer's Editor, Greenfoot, Java IDE, Mathematica, Node-RED, Scratch, IDLE 2, IDLE 3, SmartSim and Libreoffice suite.

BlueJ Java IDE, Greenfoot Java IDE are Integrated Development Environments for programming in Java. Geany Programmer's Editor is a very famous text editor that can help in writing scripts of different programming languages like PHP. Mathematica is a free tool like Matlab. It is a modern technical computing system that can be used for different applications like solving mathematical equations, data visualization, etc. Node-RED is a tool to make IoT project easily by bypassing programming and simply joining blocks. Scratch is a tool to teach Kids programming it allows them to code by joining blocks , it is similar to MIT App Inventor. IDLE is one of the most famous IDE for Coding in Python. IDLE 2 is for Python 2 and IDLE 3 is for Python 3. SmartSim is a free tool for circuit designing. Lastly LibraOffice is free alternative to MS Office. All these tools makes Raspberry Pi a great option for learning programming.

Raspberry Pi runs full Linux and can be used for teaching basics of Networking , System Programming etc.Moreover it has 40 GPIO (General Purpose Input Output) pins that can be used to interface any sort of sensor with it making it an excellent choice for IoT and automation projects. Moreover it can be used for machine learning related projects. A stripdown version of Anaconda named Miniconda exsists for Raspberry Pi. Moreover, tensorflow can also run on it. The compute power of this board is not enough to train neural networks on it but for that purpose Google's Colab can be used which is free and we only need a computer and an internet connection to use it which can be done on it.

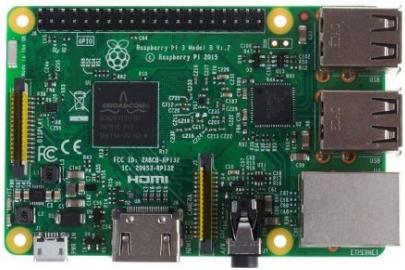

[Fig. 1] : Raspberry pi 3B SBC

### B. Raspberry pi 3B Plus

This variant improved and faster version of Raspberry pi 3b. It has the same processor but is clocked at 1.4 GHz.It has improved Ethernet which is a Gigabit Ethernet (via USB channel) . It also have dual antenna Wi-Fi which 2.4GHz and 5GHz 802.11b/g/n/ac and a better Bluetooth which is Bluetooth 4.2, Bluetooth Low Energy (BLE) [13] . It can be imported from Aliexpress along with HDMI cable, power supply, Memory card, heat sink and a case for about 55 US Dollars. Figure 2 shows Raspberry pi 3b Plus.

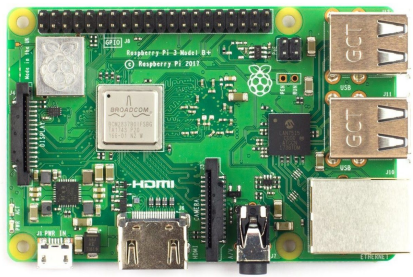

[Fig. 2] : Raspberry pi 3B Plus SBC

### C. BeagleBone Black

Just like Raspberry pi BeagleBone Black is also one of the few famous Single Board Computers. Its kit can be imported through Aliexpress for about 55 US dollars which includes the board and a USB type A wire through which it can be connected to other computers. It creates its own network and runs DHCP protocol via USB, when it is connected to other computers. It appears as a USB drive and has an introductory guide in it. This board has onboard storage of 4 GB in its C revision which is the latest one. It comes with Armgston OS (also a Linux distribution) pre-installed on it. So one just needed to plug and play. Along with that it does supports other distributions including Ubuntu. This board has Texas Intrument's AM335x 1GHz ARM Cortex-A8. Figure 3 shows Beaglebone Black.

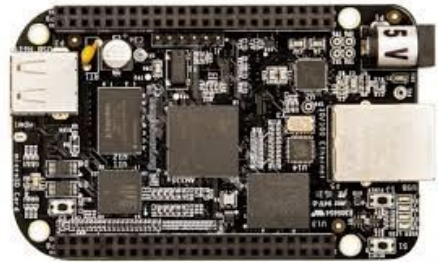

[Fig. 3] : BeagleBone Black SBC

### D. Parallela

As the name suggest this Single Board Computer is majorly designed for teaching parallel computing. It has 16-core Epiphany RISC Core SoC (System on Chip) and Zynq SoC (FPGA + ARM A9) . When it was released many called it the cheapest supercomputer available in market. Epiphany chips has a scalable array of RICS processors which can be programmed in bare metal using C and C++. Moreover major famous parallel processing frameworks are supported on it including OpenMP , OpenCL and different implementations of MPI. It is the most suitable board for someone interested in entering the field of High Performance Computing. It has 1 GB of RAM and Gigabit Ethernet Port and 48 GPIO pins. .It supports different Linux Flavours including Ubuntu. Figure 4 shows parallel SBC.

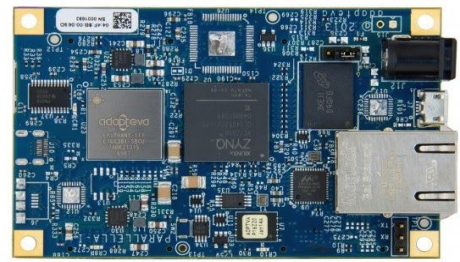

[Fig. 4] : Parallela SBC

### E. Pine 64 LTS

Pine 64 LTS is a cheaper alternative to Raspberry Pi 3B Plus. It has Allwinner A64 SoC. This chip has cortex-a53 ARM cores the same ones which are present in Raspberry pi. Allwinner is a chinese manufacturer of SoC. A64 has Mali's GPU named Mali 400 MP2. The features which is has better then Raspberry Pi includes 2 GB LPDDR3 RAM, Real Time Clock Port (RTC) and 3.5″ Barrel Power (5V 3A) Port.It was a kickstarter few years back now in 2019 its software is fear enough to make it a useful board. Like Raspberry Pi it also support many flavours of Linux including Armbian, Arch

Linux etc. Figure 5 given below shows Pine A64 LTS version.

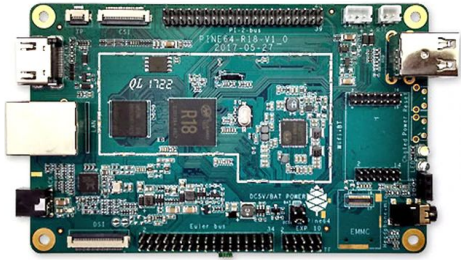

[Fig. 5] : Pine A64 LTS SBC

*F. Coral Dev Board*

AI is becoming part of every project nowadays. In recent times use of AI has come to the edge level where embedded devices have to predict the output from machine learning model to perform actions in the real world. Famous examples include object recognition and path identification; which are necessary for robots to move. The increasing use of AI has made necessary, the need to think of efficient methods for teaching it. For that purpose, we recommend to use Coral Dev Board.

This board comes with Google Edge TPU coprocessor which accelerates machine learning inferences. It is capable of performing 4 trillions operations per second. It has quad Cortex-A53, Cortex-M4F SoC as main CPU, 8 GB eMMC Flash memory and 1 GB LPDDR4 RAM. Like Raspberry Pi it also have GPIO pins and Wireless connectivity options such as WiFi and Bluetooth. It supports Python and C++ programming languages and the detailed documentation is available online for Machine learning APIs and libraries like Google TensorFlow Lite and AutoML Vision Edge. It run customized Linux named Mendel.

It is suitable for calculating inferences from pre-trained machine learning models at a very high speed which are essential in today's world to solve real world problems. With the help of online tools provided by google machine learning models can be trained online and with the Mendel Development Tool the model can be transferred to the Dev Board and user can integrate it with his program to solve real world problems. Figure 6 shows Coral Dev Board SBC.

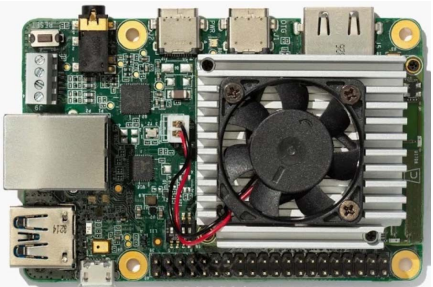

[Fig. 6] : Coral Dev Board

IV. RECOMMENDED BOARDS FOR PARTICULAR COURSES

These recommendations (table 2) are based on literature, available online resources and authors personal views.

TABLE II. RECOMMENDATIONS

| Board | Description |
|---|---|
| Raspberry Pi 3B / 3B Plus | This SBC is suitable for teaching programming languages like Python and C++, networking, system administration, parallel computer technology, IoT, computer vision, image processing and Linux OS. |
| BeagleBone Black | This SBC is a viable option for teaching C Programming and Linux OS. Being open-hardware it can also be used to teach Computer Hardware Design to Engineering students. |
| Parallela | This board is best suitable for parallel computing related courses. it includes an FPGA which makes it a viable candidate for teaching computer architecture as well. |
| Pine 64 LTS | This SBC is an affordable alternative to Raspberry Pi. Hence can be used in all cases in which Raspberry Pi can be used. |
| Coral Dev Board | Learning by doing is the best way to remember and absorb concepts. The board can be used to build robots or other AI-capable projects. It can be used for AI training. This board is the right candidate to teach AI to beginners, combining development tools and extensive documentation this board is a complete AI-Training Kit. |

V. CONCLUSION

In this article we have suggested that Cost Efficient Single Board Computers having ARM processors along with Open source software can be used for teaching; especially in developing countries like Pakistan. A comprehensive literature review has shown that people all over the world have effectively used SBCs to teach different courses. In addition, a few boards which can be used to teach certain types of courses have also been recommended.